# Small angle neutron scattering study of the step-like magnetic transformation in Pr$_{0.70}$Ca$_{0.30}$MnO$_3$

D. Saurel[1,2], Ch. Simon[1], A. Brûlet[2], A. Heinemann[3], C. Martin[1]

[1] Laboratoire CRISMAT-ENSICAEN, 6, boulevard du Maréchal Juin, 14050 Caen Cedex 04, France

[2] Laboratoire Léon Brillouin, CEA-Saclay, 91191 Gif-sur-Yvette, France

[3] Hahn Meissner Institute, Glienicker Straße 100, D-14109 Berlin, Germany

## Abstract

Small angle neutron scattering (SANS) magnetic and electrical transport measurements were performed to study a single crystal of Pr$_{0.7}$Ca$_{0.3}$MnO$_3$, a colossal magnetoresistive (CMR) material. While the magnetic field induced transformation of this phase separated compound consisting of an antiferromagnetic insulating phase (AFI) and a ferromagnetic insulating phase (FI), is continuous at high temperature (above 5K), at lower temperature a step like transformation is observed (around 5T at 2K). Macroscopic magnetization measurements and SANS indicate that this transformation occurs by the formation of mesoscopic ferromagnetic metallic (FM) domains in the AFI phase, and, eventually, in the FI phase. Although above 5K this transformation is continuous, below 5K a magnetization step marks the abrupt transition from a large scale FI/AFI phase separation to a large scale phase separation between AFI, FI and FM phases. Our results suggest that relaxation of elastic strains inherent to the coexistence of these different phases plays a crucial role in the mechanism of these transformations. The occurrence of magnetization steps could result from an intrinsic behavior of the AFI phase at low temperature.

## Introduction

Mixed valence manganese perovskites A$_{1-x}$B$_x$MnO$_3$ are widely studied owing to their colossal magnetoresistive (CMR) properties, [1] where an initial insulating phase (typically antiferromagnetic) transforms into a ferromagnetic metallic phase (FM) by application of a strong external magnetic field. Of particular interest are materials that have a step-wise magnetic field induced transformation at very low temperature (typically below 5K), and a continuous transformation at higher temperatures. For example, this behavior first was observed in Pr$_{1/2}$Ca$_{1/2}$MnO$_3$ samples that were substituted on the manganese site. [2,3,4,5] Such substitution strongly decreases the magnetic field at which a metamagnetic transition occurs, and these compounds also present several magnetization steps. For CMR Pr$_{1-x}$Ca$_x$MnO$_3$ samples with x values smaller than 0.5,[6,7,8,9] and which are not doped on the Mn site of the manganite, a metastable AFI phase is observed (e.g. Pr$_{5/8}$Ca$_{3/8}$MnO$_3$ samples [9]).

In previous reports, the aforementioned step-like transitions have been compared to the martensitic transitions displayed by shape memory alloys.[10] This would suggest that strains between coexisting phases (initial and field induced ones) play an important role in the occurrence of the magnetization steps, and that the transformation is strongly inhomogeneous. In a previous paper,[5] we presented a detailed study of the step-like magnetic transformation of a polycrystalline sample of Pr$_{0.5}$Ca$_{0.5}$Mn$_{0.97}$Ga$_{0.03}$O$_3$, by magnetization, neutron diffraction and small angle neutron scattering (SANS). Because the latter technique is well adapted to study magnetic inhomogeneities in the range of 0.5 to 50 nm, the domain size of the co-existing F and AF phases which present a nice contrast for neutron scattering, can be estimated. Indeed we have shown by neutron diffraction that Pr$_{0.5}$Ca$_{0.5}$Mn$_{0.97}$Ga$_{0.03}$O$_3$ is phase separated between two distinct AF crystallographic phases at low temperature.[5] Evolution of the SANS intensity versus magnetic field was typical of inhomogeneous transformations of initial AF phases into a FM one, in agreement with the martensitic scenario. However, the presence of a large scattering, owing to grain boundaries in this polycrystalline sample, prevented a more quantitative analysis. Moreover, in $Pr_{0.5}Ca_{0.5}Mn_{0.97}Ga_{0.03}O_3$, the metastability of the initial AF phases is induced by the doping on the Mn sites of the perovskite by Ga ions. In this case, it is difficult to claim that the origin of these step-like transformations is, or is not, an intrinsic property of the AF phases, as it is proposed by some authors[6,9].

In this paper, we report a study of a $Pr_{0.7}Ca_{0.3}MnO_3$ single crystal without any substitution on the Mn site. This compound is one of the $Pr_{1-x}Ca_xMnO_3$ series, widely studied for their remarkable CMR properties in the range $0.25 < x < 0.40$.[11,12,13] Furthermore, in absence of magnetic field, this compound is known to present a phase separation between two distinct crystallographic phases, both insulating, FI and AFI.[14,15] The application of the magnetic field leads to a metallic ferromagnetic phase (FM). We have recently shown by SANS that in the temperature range 10-30K where the magnetic transformation is continuous, both AFI and FI phases are transformed into FM phase by magnetic field.[16] Here we study in detail this compound at lower temperature where the magnetic transformation is step-like. SANS and magnetization curves were analyzed and compared to their variations at low (10K-50K) and very low (2K-5K) temperatures to go further into the understanding of FI and AFI phases transformations under magnetic field.

## Experimental

A several centimeters long single crystal was grown using the floating-zone method with a feeding rod of nominal composition $Pr_{0.7}Ca_{0.3}MnO_3$. Two small samples were cut out from the crystal in order to perform the different measurements: a cube of about 10 $mm^3$ for the magnetization and transport measurements, and a parallelepiped of about 40 $mm^3$ for the SANS measurements. Prior to characterizations, the purity of the sample was checked by energy dispersive spectroscopy, X-ray, electron and neutron diffraction.

Measurements of magnetization as a function of the magnetic field up to 9T were performed using the standard extraction magnetometry on a Quantum Design PPMS magnetometer. Both samples show similar magnetization curves up to 5.5T. Conductivity data were acquired via the standard four-probe technique. Data were recorded by increasing the magnetic field unless specified on the figures, in the zero field-cooled procedure. The cooling rate was 10K/mn and the field was ramped at 250 Oe/sec.

SANS was performed on the "PAXY" spectrometer of the LLB at the Orphée reactor and on the "V4" spectrometer of BENSC in Hahn-Meitner Institute (HMI). Different experimental configurations were used to reach a range of scattering vector Q from 0.05 $nm^{-1}$ to 2 $nm^{-1}$ with a large overlap between each range. The sample was introduced in a cryostat with aluminum windows. The data treatment method used to extract quantitative data in absolute units ($cm^{-1}$) has been previously reported.[17] No precise orientation of the crystal was chosen. Without magnetic field, the SANS pattern is isotropic, which reveals that, in this Q range, both nuclear and magnetic scattering are isotropic. In presence of magnetic field, the anisotropic signal observed is due to the orientation of magnetic moments. The scattering can thus be separated into two contributions. The first one, $I_A(Q)$, is isotropic and is the sum of the nuclear scattering and the scattering by the magnetism not oriented by the magnetic field. The second one, $I_B(Q)\sin^2\alpha$, is a pure magnetic anisotropic scattering due to the magnetism oriented by the magnetic field :[18]

$$I(Q,\alpha) = I_A(Q) + I_B(Q)\sin^2\alpha$$
(1)

where $\alpha$ is the angle between the scattering vector Q and the direction of the applied magnetic field. In this paper, we analyze the magnetic field evolution of the pure magnetic signal $I_B$, extracted using the BERSANS software (from HMI) which fits I(Q,$\sin^2\alpha$) as a linear function.

### Results and discussion

The magnetization as a function of external magnetic field is represented in figure 1 with different scales, for temperatures ranging from 2K to 30K. At 30K (figure 1a), from zero field up to 1.5T, the curvature is attributed classically to the orientation of the ferromagnetic domains of the initial FI phase.[16,19] Between 1.5 and 3T, the magnetization curve displays a constant slope, indicating that there is no magnetic transformation, and can be attributed to the susceptibility of the AFI phase. The projection of this linear part to B=0, divided by the saturated magnetization at high field (9T), gives the fraction of the initial FI phase, and is estimated to be 31% +/- 1% at 30K. Above 3T, the magnetization increases rapidly, owing to the continuous transformation of the AFI phase into a FM one. Above 6T, the magnetization saturates at $4.2\mu_B$, which corresponds to the complete transformation into the FM phase.

Magnetization curves shown in figure 1 qualitatively follow the same scenario. In contrast, below 5K, we observe a discontinuity between 5T and 5.25T in the transformation of the AFI phase. Indeed, the curves show a linear regime at low field. Just beyond this regime and around 4-4.5T, the magnetization increases slightly more rapidly, , due to the transformation of the AFI phase. Then between 5T and 5.25T, an abrupt jump of the magnetization is observed. At higher fields, the magnetization curves show a plateau, and the continuous transformation starts again, at 6-6.5T and reaches saturation around 8-8.5T. The characteristic magnetic fields deduced from these magnetization curves have been summarized in the temperature field phase diagram (figure 2). The open circles represent the values of the magnetic field at which this transformation begins and ends. The discontinuity observed between 5T and 5.25T at 2-5K is represented by the black squares. The low field zone is a region of phase separation between FI and AFI phases.[14] The high field zone, when the saturation of the magnetization is reached, corresponds to a fully FM system. The gray zone of the phase diagram corresponds to continuous transformation of the different phase fractions as the field is applied. The low temperature white zone in the middle corresponds to the plateau of constant magnetization observed when increasing the magnetic field just after the magnetization jumps. In this domain, the three phases coexist and their phase fractions are constant.

Since the evolution of the resistance versus the magnetic field (inset of figure 1c) shows a discontinuous insulator to metal transition between 5 and 5.25T, we conclude that this intermediate state is a conducting one. We observe that once the sample has reached a metallic state and when the magnetic field is switched off, the sample remains in this final state (figure 1a). To recover the initial zero field FI/AFI state, the sample must be annealed above its orbital ordering transition temperature $T_{OO}$ ~180K [14].

To extract more quantitative information from the magnetization curves of figure 1, we have used the following expression to represent the magnetization:[9,16]

$$M = \phi_F \, M_F + (1 - \phi_F) \chi_{AF} B$$
(2)

where $\phi_F$ is the amount of F phase (sum of the initial amount of FI and of the field induced FM) and $1 - \phi_F$ the amount of the AFI phase. $M_F$ is the local magnetization of the F phase, estimated to be equal to the saturated magnetization (i.e. $4.2 \, \mu_B$ /f.u.) and $\chi_{AF}$ is the susceptibility of the AFI phase. The amount of total F phase $\phi_F$ extracted from the magnetization curves using equation (2) has been plotted in figure 3a at the typical temperatures of 2K and 30K. These plots allow the transformation of the AFI phase to be monitored which is useful in analyzing the evolution of the SANS curves.

Typical examples of pure anisotropic magnetic scattering curves $I_B(Q)$ are plotted in figure 4, at 2K (2T, 5.9T) and at 30K (2T, 4T). 5.9T (and 4T) are field values at 2K (and 30K), respectively, where part of the initial AF phase has been transformed into the FM phase (figure 3a). Curves obtained at 2K and 30K are quite similar and two features at two different Q ranges can be distinguished.

At low Q values, we observe a $Q^{-4}$ component, which is characteristic of sharp interfaces of scattering objects :[20]

$$I_B(Q) = 2\pi(\Delta\rho)^2 \, S/V \, Q^{-4}$$
(3)

where S/V is the specific area of the scattering objects and $(\Delta\rho)^2$ the contrast, here between the F and the AF phases. The evolution of $Q^4 I_B$, which is directly proportional to the total interface area S, is illustrated in figure 3c. At 2K and 30K, $Q^4 I_B$ is zero at low magnetic field. At 30K, upon increasing the field, a $Q^{-4}$ scattering appears when the transformation of the AF phase observed on the magnetization curve, (figure 3a). We have demonstrated that the peculiar "triangular" shape of the $Q^4 I_B$ curve at 30K (figure 3c) was typical of a nucleation, growth and collapse of FM regions in the largest AFI phase.[16] The total amount of interface S increases with nucleation and growth, reaches a maximum when 50% of the initial AFI phase is transformed into the FM phase and then decreases with the amount of AFI phase. At 2 K, a $Q^{-4}$ component also appears but at 6T, while it was not observed below the step (2T and 4T). This means that the amount of interfaces S has increased. If we assume, as at 30K, a complete contrast between the AFI and FI regions, using the FM+FI phase fraction obtained from magnetization curves (equation (2)), we deduce an average size of FM domains of 1.3 μm, i.e. of the same order of magnitude than the initial size of the AFI regions estimated at 30K. Thus, at 2K and 6T, we are in presence of a large scale (micrometric) phase separation between AFI and FM phases. At 2K, we have no data between 4T and 6T, but we may assume that the continuous transformation represented by the gray zone on figure 2a is inhomogeneous, similar to the one observed at 30K. In other words, we think that at low temperature (below 5K), the transformation starts by continuous nucleation of mesoscopic FM clusters in the AFI phase, and between 5 and 5.25T, an abrupt reorganization of these two phases into a large-scale phase separated system occurs.

At larger Q values, the $I_B(Q)$ curves, shown in figure 4, obey to a $Q^{-2}$ power law attributed to non-magnetic nanometric inhomogeneities in the FI phase.[16] As previously reported,[16] the shape of this signal does not change with the temperature (in the range 2K-30K) or magnetic field, whereas its intensity decreases when the magnetic transformation proceeds. We simply attribute the decrease of this $Q^{-2}$ signal as owing to a decrease of the amount of FI phase.[16] The intensity at a high Q value of this SANS component thus provides a good measurement of the FI phase fraction, $\phi_{FI}$, as shown on figure 3b, the variation of $I_B$ at Q=1nm$^{-1}$ versus the magnetic field. Combining magnetic and SANS measurements, albeit is possible to determine the magnetic field dependence of percentages of FI, AFI and FM phases (figure 3 at 2K, 10K and 30K). It appears that at both temperatures, the FI phase starts to transform about 0.5T after the AFI one. The transformation of the AFI phase is clearly due to the magnetic energy between the magnetic moments and the magnetic field. In contrast, we cannot use the same argument for the transformation of the FI phase into FM phase. We propose that the interface between FM and FI is more strained than the FI to AFI interface. In order to reduce its lattice elastic energy, the system decreases the amount of FI phase as the FM phase is growing. P. G. Radaelli et al.[15] have suggested previously that the appearance of FM clusters can be attributed to strains in the lattice. As shown in figure 5, at 2K and 6T, amounts of both FI and AFI phases have decreased significantly compared to 2T and 4T. Since the loss of FI phase is larger than the decrease of AFI phase observed between 4.5T and 5T (1.3%), we can assume that part of the FI phase has been also transformed at the magnetization step. At the conclusion, around 5T, the system has transformed abruptly from a large scale 35%/64% FI/AFI phase separation, with AFI containing 1% of small FM clusters, to a large scale FI/AFI/FM phase separation, with phase fractions 27%/8%/65%, respectively. Such phase transformations occurring at the magnetization step correspond to a huge change in strain fields in the system. Here also, as at higher temperature, the transformation of the FI phase is probably due to the need to reduce the elastic energy of the system. It would stop when the equilibrium minimizing the elastic energy is reached. This phenomenon corresponds to the plateau observed above 5.25T in the magnetization curves in the range 2K-5K. The typical size of the large FI/FM/AFI phase separation corresponding to this plateau (about 1μm) is similar to the domain size obtained at 30K, 5T, when the continuous transformation is well advanced. Thus we propose that the transformation has the same phenomenology although continuous in one case and discontinuous in the other case: when the amount of FM clusters in the AFI phase induced a critical amount of strains, the system reorganizes by displacing the AFI/FI interfaces to provide place for larger FM domains. This reorganization induced by strains is analogous with the martensitic transformations previously reported.[10]

Although our results show that strains take an important place in the mechanism of these transformations, it is much more difficult to summarize the origin of the magnetization steps. On the inset of figure 1a, we have plotted the temperature dependence of the percentage of initial FI phase deduced from the regime of constant slope of the magnetization versus magnetic field. Above 5K, we observe a clear decrease in this percentage, which can probably be attributed to minimizing the strain field between the two coexisting initial FI and AFI phases. Nevertheless, below 5K, this percentage is roughly constant, suggesting that at very low temperature the strain field cannot be minimized lower than at higher temperature, and is released at the step. Moreover, we have shown that this step occurs at the same magnetic field independent of the temperature up to 5K (black squares on figure 2a). Although the magnetic field at which the transformation starts varies with temperature (open circle of figure 2a). All these results reveal different stabilities of initial FI and AFI phases versus magnetic field and thermal agitation below and above 5K, which are related to the occurrence of steps. Because the system is out of equilibrium during the step transformation, the study of the step dynamics, which is not possible by SANS, will be the focus of future studies.

## Conclusion

$Pr_{0.70}Ca_{0.30}MnO_3$ presents three different magnetic phases at low temperature (FI, AFI, FM), with quantities depending on the applied magnetic field. Magnetization measurements give the evolution of the amount of AFI phase. Determination of the quantities of the FI phase is possible by SANS, which takes advantage of the presence of a large magnetic SANS signal arising from the FI phase. We have thus determined the B-T phase diagram of this compound. Below 5K, the magnetic transformation is step-like whereas it is continuous at higher temperatures. Our study allows describing in detail the transformation phenomena from 2K to 30K. We conclude that in both cases, the scenario is consistent with the martensitic transitions. The magnetic transformation starts by nucleation and collapse of mesoscopic FM clusters in the AFI phase. At a certain magnetic field, the transformation is transmitted to the FI phase. Indeed, a noticeable difference of stability of the AFI phase above and below 5K is highlighted, and the appearance of an intermediate metastable state at low temperature in which the system falls over at the step. An energy barrier limits the transformation of the AFI phase during the cooling, which suggests that the origin of magnetization steps is more an intrinsic behavior of the AFI phase than a consequence of phase separation.

## *Acknowledgments*

We acknowledge L. Hervé for samples preparation, V. Hardy, A. Maignan, M. Hervieu and R. Retoux for the numerous scientific discussions and F. Ott for his help during the experiments. C. Sheets spend a lot of time reading carefully the manuscript. D. Saurel was supported from CEA and "Région Basse Normandie". Partial support from the European Community under the STREP research project CoMePhS (No. 517039) is also acknowledged.

## *Figure captions*

Figure 1: Evolution of magnetization versus applied magnetic field below 30K (zero field cooled) of the $Pr_{0.7}Ca_{0.3}MnO_3$ crystal.

a- M(B) curves at 30K and 5K (1) and (2). At 5K, the field decrease magnetization (2) and the reproducibility of the measurements after room temperature annealing are also shown. Inset: temperature dependence of the FI phase fraction determined from magnetization curves.

b- Zoom in the region of the magnetization step at 2, 3, 5, 10 and 30K.

c- Zoom just before the magnetization step at 2K and 5K. The departure from linearity appears before the magnetization step. Inset: field evolution of the resistance at 2K.

Figure 2:

a- Temperature - field phase diagram of $Pr_{0.7}Ca_{0.3}MnO_3$ from magnetization (open circles and black squares) and SANS (open triangles). The gray domain is where FM coexists with the insulating phases and its fraction varies continuously with magnetic field. The black squares mark the magnetization steps. The triangles mark the beginning of the transformation of the FI phase.

b to f: schematic pictures representing the evolution of the magnetic microstructure.

Figure 3:

a) Evolution of the amount of ferromagnetic phase (initial FI + field-induced FM) deduced from the magnetization curves of figure 1.

b) Evolution of the $I_B$ intensity scattered at $Q=1nm^{-1}$ by non-magnetic nanometric inhomogeneities of the FI phase.

c) Evolution of the $Q^4 I_B$ component, due to the scattering by FM regions in the AFI phase, deduced from the $I_B(Q)$ for $Q \leq 0.1nm^{-1}$ (figure 4).

Figure 4: Purely magnetic SANS curves $I_B(Q)$ of the $Pr_{0.7}Ca_{0.3}MnO_3$ crystal at low magnetic field (2T) at 2K (a) and 30K (b) 5.9T at 2K (c) and 4T at 30K (d). Lines are guide for eyes.

Figure 5: Evolution of the FI, AFI and FM phase fractions deduced from SANS and magnetization data at 30K, 10K and 2K. AFI fraction (hollow circles) is obtained from magnetization measurements, FI percentage (black squares) from the $Q^{-2}$ SANS signal. Lines are guide for eyes.

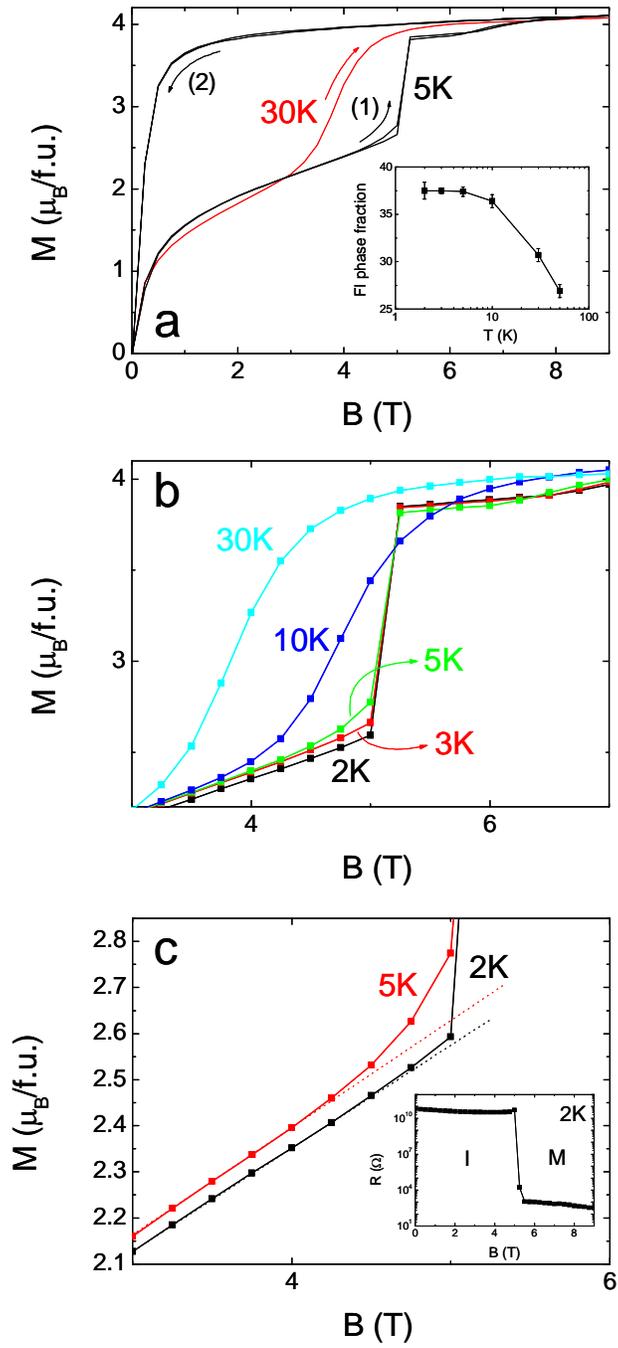

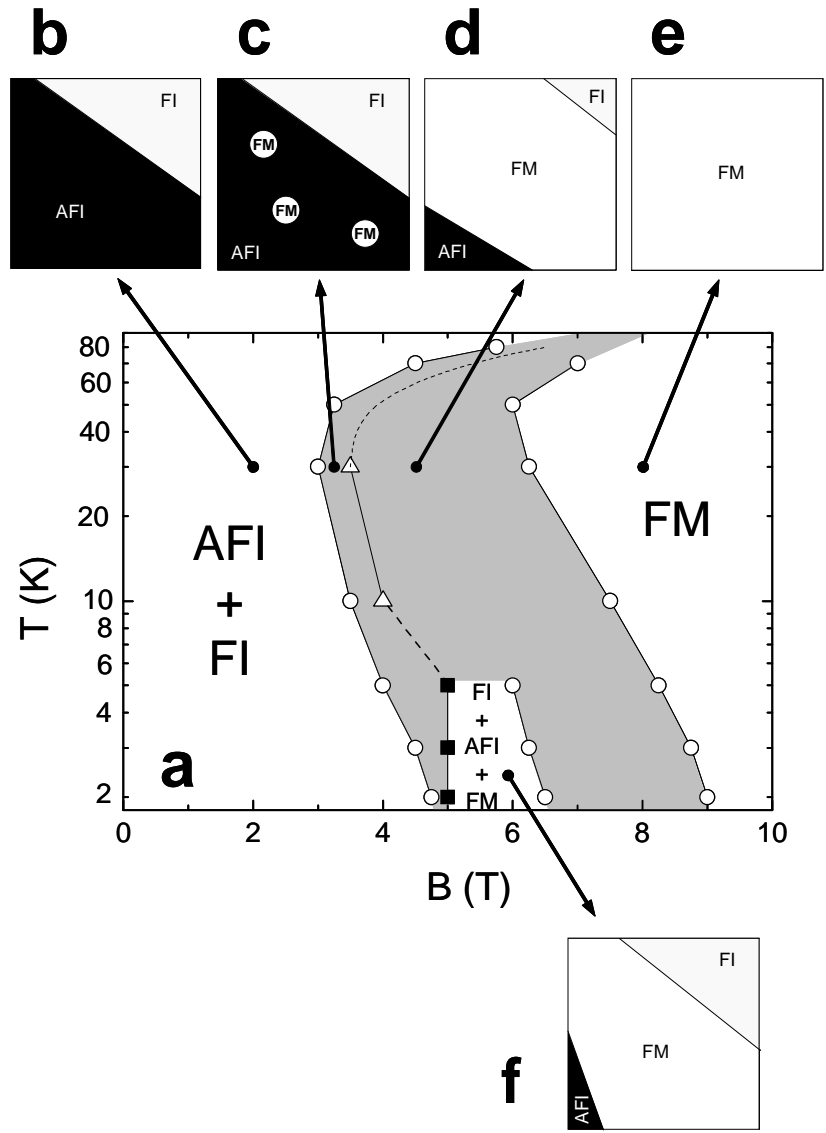

*Figure 2*

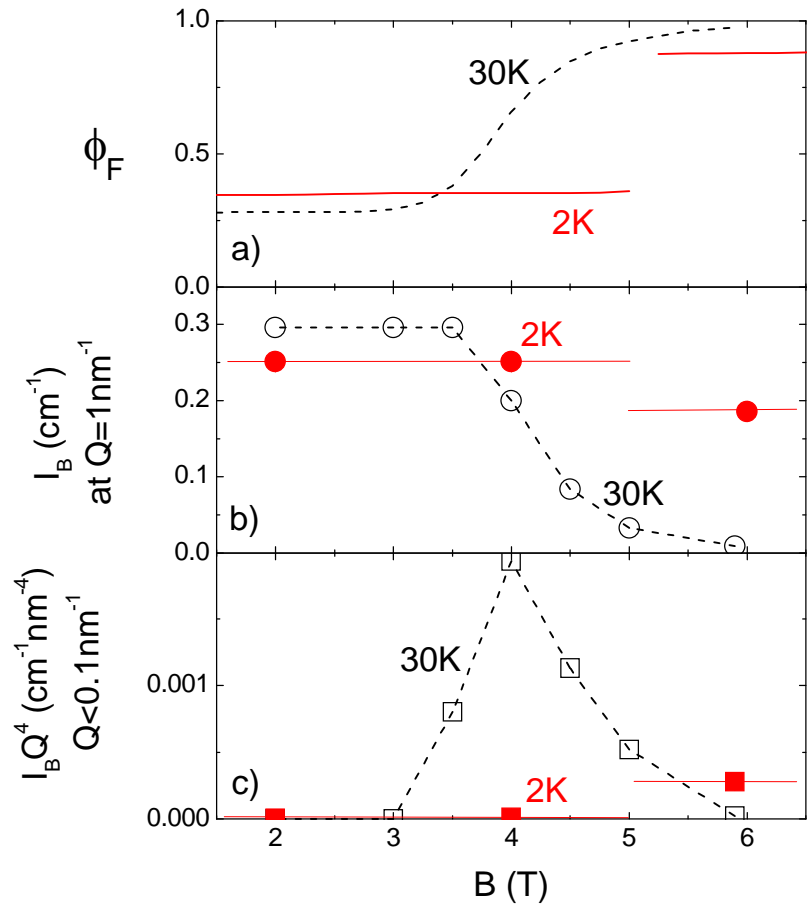

*Figure 3*

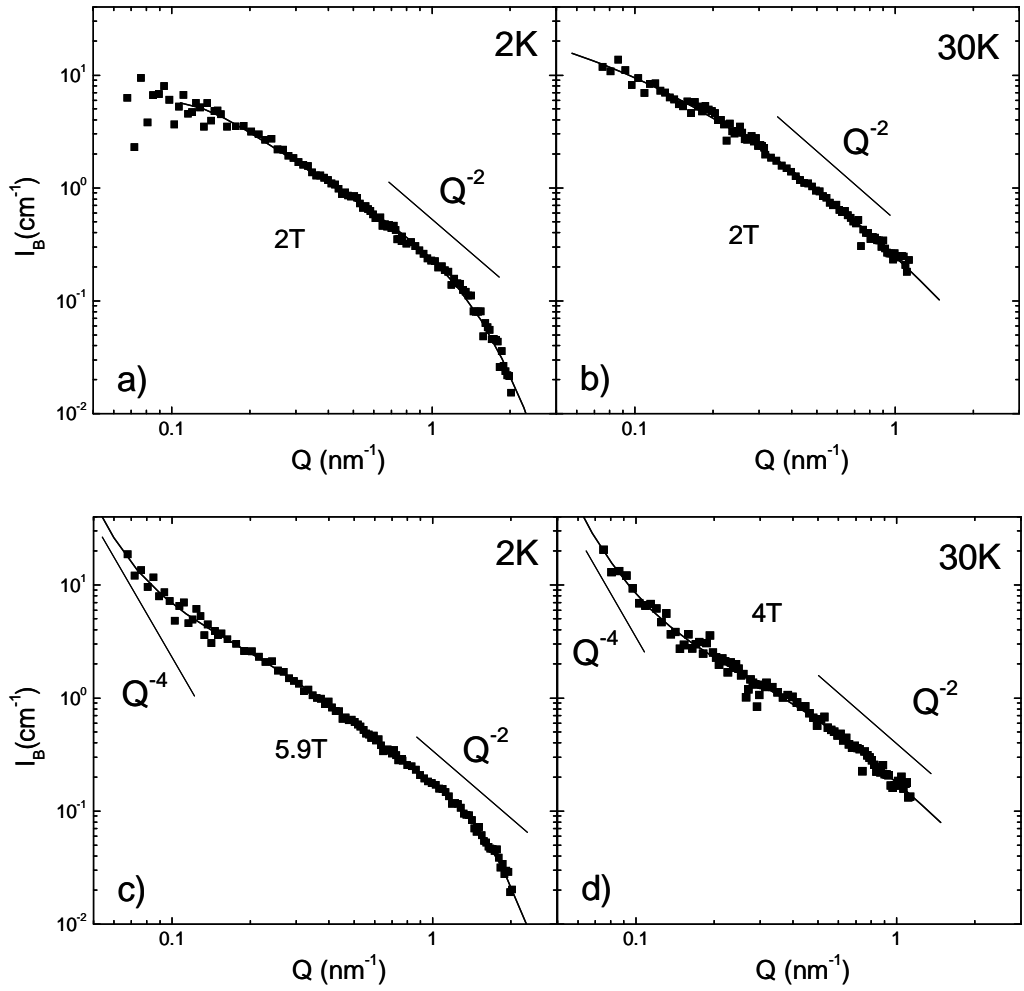

*Figure 4*

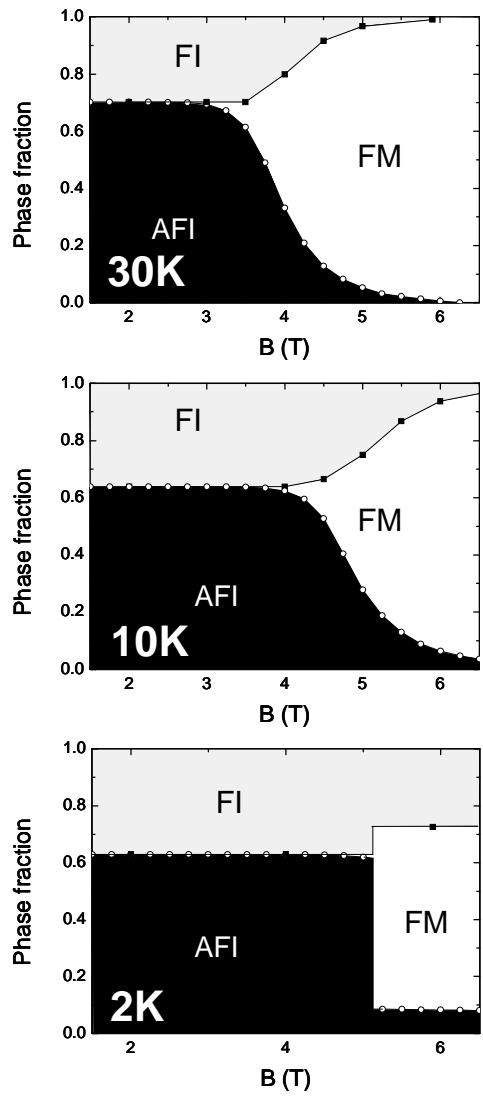

*Figure 5*